\documentclass[aps,twocolumn,superscriptaddress,a4paper,floatfix]{revtex4-2}
\usepackage[latin1]{inputenc}
\usepackage{graphicx}
\usepackage{amsmath,amssymb}
\usepackage{hyperref}

\newcommand \beq{\begin{eqnarray}}
\newcommand \eeq{\end{eqnarray}}

\newcommand \mk{\mathbf{k}}

\begin{document}

\title{Particle current, noise, and counting statistics of quantum transport in the presence of a single-particle loss}

\author{Shun Uchino}
\affiliation{Advanced Science Research Center, Japan Atomic Energy Agency, Tokai 319-1195, Japan}


\begin{abstract}
How dissipation affects transport is an important theme in quantum science.
Here we theoretically investigate an impact of a single-particle loss in mesoscopic transport, which has been an issue in experiments of ultracold atomic gases.
By explicitly analyzing quantum point contact and quantum dot systems,
we obtain a cumulant generating function on the particle current whose formal expression turns out to be common to two systems.
In terms of this generating function, 
behaviors of  average current, particle loss rate, and noises in presences of losses introduced in conduction channels are exemplified for free fermions.
It is shown that the current noise contains the component proportional to the particle loss rate, which may be measurable in experiments.

\end{abstract}


\maketitle

\section{Introduction}
Understanding impacts of dissipation has been a key issue in quantum transport~\cite{nazarov}.
The renowned example is dissipation in Josephson junctions~\cite{schon1990quantum}.
There, an Ohmic current is known to be present even at low temperatures at which
a quasiparticle current is expected to be suppressed due to a superconducting gap.
In the classic paper, Caldeira and Leggett suggested that dephasing invoking the Ohmic current comes out
due to a coupling to a reservoir that is modeled by a bath of harmonic oscillators~\cite{PhysRevLett.46.211}.
Dissipation associated with dephasing is also the matter
in controllable mesoscopic transport with condensed matter systems. 
For instance,  an occupation of a quantum dot is known to affect a current of a quantum point contact (QPC) in a coupled system of quantum dot and QPC~\cite{sukhorukov2007conditional},
which can be understood by means of a quantum master equation with dephasing~\cite{wiseman}.

In addition to condensed matter systems, ultracold atomic gases being artificial quantum systems trapped in a vacuum
provide an alternative route to investigate dissipation in quantum transport~\cite{ott2016single}.
By using focused electron beam~\cite{PhysRevLett.116.235302,mullers2018coherent}, photoassociation~\cite{tomita2017observation} and optical tweezer~\cite{PhysRevA.100.053605}, 
ultracold atomic gases naturally enable us to introduce atom losses as well-controlled dissipation.

In particular,  the ETH group has recently succeeded in manipulating mesoscopic systems such as a two-terminal QPC system with a quantum gas microscope~\cite{krinner2017}
and analyzed an effect of a single-particle loss in mesoscopic transport.
In Ref.~\cite{PhysRevA.100.053605}, a local single-particle loss imprinted in the conduction channel region of the QPC system is found to
lead to a reduction of the conductance~\cite{PhysRevA.100.053605}.
Later on, such a reduction has been explained in terms of a quantum master equation with non-unitary time evolution~\cite{PhysRevLett.129.056802,PhysRevA.106.053320,PhysRevResearch.5.013195}.
More recently, an effect of the single-particle loss has been investigated with superfluid reservoirs,
where nonlinearity originating from multiple-Andreev reflections is weakened by dissipation~\cite{PhysRevLett.130.200404,visuri2023dc}.

\begin{figure}[htbp]
\centering
\includegraphics[width=5.5cm]{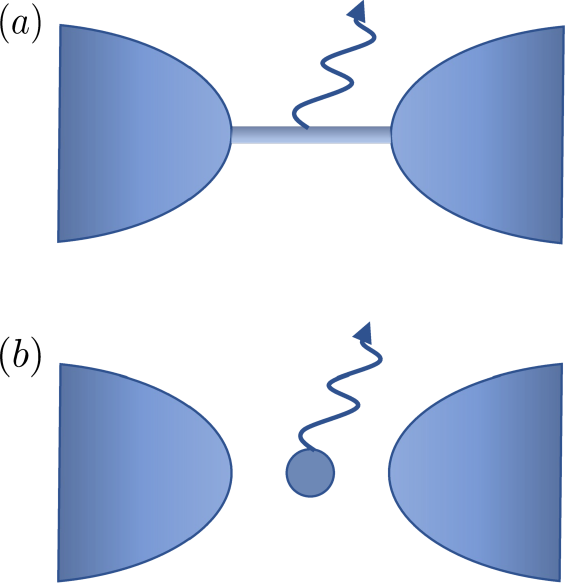}
\caption{\label{fig1}
Dissipative two-terminal mesoscopic systems discussed in this work: (a) quantum point contact and (b) quantum dot.
Building on mesoscopic transport experiments of ultracold atomic gases, we focus on the situation that a single-particle loss occurs solely in the conduction channel. } 
\end{figure}
Until now,  mesoscopic transport experiments with ultracold atomic gases  focus on an average current.
We note that this is in contrast to condensed-matter experiments in which current fluctuations such as noises have also been measured~\cite{blanter2000}.
For example, noise measurements allow us to observe an effective charge of quantum many-body systems including the fractional charge of a fractional quantum-Hall state~\cite{PhysRevLett.79.2526,de1998direct},
and transport relations associated with the fluctuation theorem in nonequilibrium statistical mechanics~\cite{PhysRevLett.104.080602}.
Although  in experiments of ultracold atomic gases current fluctuation effects have been masked by 
relatively large experimental errors due to destructive measurements, near future experiments including a non-destructive measurement~\cite{PhysRevA.98.063619} may overcome the difficulty.

Based on general interest of current fluctuations in mesoscopic systems and future perspective of experiments with ultracold atomic gases,
here we give a theoretical formulation of a lossy two-terminal transport system that incorporates statistics of current fluctuations.
Motivated by mesoscopic transport experiments with ultracold atomic gases,  we examine prototypical two-terminal systems, i.e., QPC (Fig.~\ref{fig1}~(a)) and quantum dot systems (Fig.~\ref{fig1}~(b)),
where a single-particle loss is present in a conduction channel region. 
It is shown that a formal expression of a cumulant generating function on the particle current  is  common to two systems
and so are the current fluctuations.
In particular, we uncover that the current noise and the noise of the particle loss rate contain components proportional to the particle loss rate,
implying that the noises become the same order as the average quantity and thus may be measured in experiments.
Despite the similarity on the generating function, frequency dependences on channel transmittance and loss probability are 
different between two systems, which leads to a significant difference in quantum transport.
In order to exemplify it, we examine the free fermion case,
although the formulation applied regardless of quantum statistics is given below.
We demonstrate that compared to the average current, the noise becomes a sensitive quantity that reflects on a frequency dependence of a loss probability.

This paper is organized as follows.
Section II gives a three-terminal Landauer-B\"{u}ttiker formulation for 
mesoscopic transport in the presence of a single-particle loss, and discusses generic expressions of the average current,
particle loss rate, and current noises.
In Sec. III, an explicit form of the $S$-matrix in the QPC is
given and applied to discuss the transport properties based on counting statistics of the current.
Section IV discusses the case of the quantum dot system.
To this end, we adopt a formulation based on the Keldysh quantum field theory approach.
In Sec, V, we wish to summarize our findings and mention an outlook.

\section{Three-terminal Landauer-B\"{u}ttiker formulation}
 We first discuss a three-terminal Landauer-B\"{u}ttiker
formalism~\cite{PhysRevB.46.12485,PhysRevB.95.085411} for single-channel quantum transport in the presence of a single-particle loss. 
The essential idea of this approach is that an additional terminal appended in the two-terminal system serves a function
of the vacuum absorbing particles in the conduction channel~\cite{PhysRevA.106.053320}.
This section focuses on generic forms of average quantities and current noise, rather than specific behaviors that depend on the conduction channel in details.

In the Landauer-B\"{u}ttiker formulation, we consider the particle current operator in each reservoir given by
\beq
\hat{I}_j(\tau)=\int \frac{d\omega}{2\pi}\int d\omega'e^{i(\omega-\omega')\tau}[\hat{a}^{\dagger}_j(\omega)\hat{a}_j(\omega')-\hat{b}^{\dagger}_j(\omega)\hat{b}_j(\omega')],
\eeq
where $j=1,2,3$ is the reservoir index, and $\hat{a}$ and $\hat{b}$ are annihilation operators for the incoming and outgoing states, respectively~\cite{blanter2000}.
The scattering process between the incoming and outgoing states is expressed as~\cite{PhysRevB.46.12485,PhysRevB.95.085411}
\beq
\begin{pmatrix}
\hat{b}_1\\
\hat{b}_2\\
\hat{b}_3
\end{pmatrix}=
\begin{pmatrix}
r & t' & l_1\\
t & r' & l_2\\
l_1' & l_2' & r_3
\end{pmatrix}
\begin{pmatrix}
\hat{a}_1\\
\hat{a}_2\\
\hat{a}_3
\end{pmatrix}.
\eeq
Here, the above $3\times3$ matrix connecting $\hat{a}$ with $\hat{b}$ is called $S$-matrix and
satisfies the unitary condition, i.e.,
\beq
\hat{s}\hat{s}^{\dagger}=\hat{1}.
\eeq
The above leads to the following relationships that are useful in the subsequent analyses:
\beq
|r|^2+|t'|^2+|l_1|^2=1,\\
|r'|^2+|t|^2+|l_2|^2=1,\\
|r_3|^2+|l_1'|^2+|l_2'|^2=1,\\
r't'^*+tr'^*+l_1l_2^*=0.
\eeq
In addition, by using the $S$-matrix, the current operators can solely be expressed with the incoming operators as follows:
\beq
\hat{I}_k(\tau)=\sum_{i,j=1,2,3}\int\frac{d\omega}{2\pi}\int d\omega' e^{i(\omega-\omega')\tau}\hat{a}_i^{\dagger}(\omega)A^k_{ij}(\omega,\omega') \hat{a}_j(\omega'),\nonumber\\ 
\eeq
where
\beq
A^k_{ij}(\omega,\omega')=\delta_{i,k}\delta_{j,k}-s^{*}_{ki}(\omega)s_{kj}(\omega').
\eeq
In the Landauer-B\"{u}ttiker approach, the averages on the incoming operators are assumed to satisfy
\beq
\langle \hat{a}^{\dagger}_i(\omega)\hat{a}_{j}(\omega')\rangle=\delta_{i,j}\delta(\omega-\omega')n_i(\omega),
\eeq
where
\beq
n_i(\omega)=\frac{1}{e^{(\omega-\mu_i)/T}\mp1}
\eeq
is the distribution function with the chemical potential in reservoir $i$ $\mu_i$ and temperature $T$.
It follows that minus (plus) sign in $n_i$ corresponds to bosons (fermions).

Based on the generic discussions above for a three-terminal Landauer-B\"{u}ttiker system,
we now consider a connection between three-terminal and  lossy two-terminal systems.
As shown in Ref.~\cite{PhysRevA.106.053320}, the three-terminal Landauer-B\"{u}ttiker system reduces to
two-terminal system in the presence of a single-particle loss, provided that the following conditions are satisfied:
\beq
&&{\cal T}(\omega)=|t|^2=|t'|^2,\label{eq:transmittance} \\
&&{\cal L}(\omega)=|l_1|^2=|l_2|^2=|l'_1|^2=|l_2'|^2,\label{eq:loss-prob}\\
&&\int d\omega{\cal L}(\omega)n_3(\omega)=0,\label{eq:vacuum}
\eeq
where ${\cal T}$ and ${\cal L}$ are channel transmittance and loss probability, respectively.
If the above conditions hold, the average particle current from left to right is rewritten as
\beq
I=\frac{\langle\hat{I}_1\rangle -\langle\hat{I}_2\rangle }{2}
=\int_{-\infty}^{\infty}\frac{d\omega}{2\pi}\Big[{\cal T}(\omega)+\frac{{\cal L}(\omega)}{2}\Big][n_1(\omega)-n_2(\omega)].\nonumber\\
\eeq
In addition, the particle loss rate can be given as follows:
\beq
-\dot{N}=\langle\hat{I}_1\rangle+\langle\hat{I}_2\rangle=\int_{\infty}^{\infty}\frac{d\omega}{2\pi}{\cal L}(\omega)[n_1(\omega)+n_2(\omega)].
\eeq
The expressions above are indeed consistent with results of Refs.~\cite{PhysRevA.100.053605,PhysRevA.106.053320}.

We next extend our analysis to the power spectral density of the current or simply current noise.
This quantity is defined as
\beq
S_{ij}(\omega)=\frac{1}{2}\int d\tau e^{i\omega\tau}\langle \delta \hat{I}_i(\tau)\delta\hat{I}_j(0)+\delta\hat{I}_j(0)\delta\hat{I}_i(\tau)\rangle\nonumber\\
=\int d\tau e^{i\omega\tau}\Big[ \frac{\langle \hat{I}_i(\tau)\hat{I}_j(0)+\hat{I}_j(0)\hat{I}_i(\tau)\rangle}{2}- \langle \hat{I}_i\rangle\langle \hat{I}_j\rangle\Big],
\eeq
where
\beq
\delta\hat{I}_i(\tau)=\hat{I}_i(\tau)-\langle \hat{I}_i\rangle.
\eeq
In terms of $S_{ij}$, the current noise between 1 and 2 is expressed as
\beq
S_{I}(\omega)=\frac{S_{11}(\omega)+S_{22}(\omega)-S_{12}(\omega)-S_{21}(\omega)}{4}.
\eeq
In addition, the noise of the particle loss rate is given by
\beq
S_{N}(\omega)=S_{11}(\omega)+S_{22}(\omega)+S_{12}(\omega)+S_{21}(\omega).
\eeq
To calculate $S_{ij}$, we use
\begin{flalign}
\langle \hat{a}_i^{\dagger}(\omega_1)\hat{a}_j(\omega_2)\hat{a}_k^{\dagger}(\omega_3)\hat{a}_l(\omega_4) \rangle-
\langle \hat{a}_i^{\dagger}(\omega_1)\hat{a}_j(\omega_2)\rangle\langle\hat{a}_k^{\dagger}(\omega_3)\hat{a}_l(\omega_4) \rangle\nonumber\\
=\delta_{i,l}\delta_{j,k}\delta(\omega_1-\omega_4)\delta(\omega_2-\omega_3)n_i(\omega_1)[1\pm n_j(\omega_2)],\nonumber\\
\end{flalign}
where the upper (lower) sign corresponds to Bose (Fermi) statistics.
We then obtain
\begin{flalign}
S_{ij}(\omega)
=\sum_{\alpha,\beta=1,2,3}\int \frac{d\omega_1}{4\pi}A^{i}_{\alpha\beta}(\omega_1,\omega_1+\omega)A^{i}_{\beta\alpha}(\omega_1+\omega,\omega_1)\nonumber\\
\times B_{\alpha\beta}(\omega_1,\omega_1+\omega),
\label{eq:generic-noise}
\end{flalign}
where 
\beq
B_{\alpha\beta}(\omega_1,\omega_2)=n_{\alpha}(\omega_1)+n_{\beta}(\omega_2)\pm2n_{\alpha}(\omega_1)n_{\beta}(\omega_2).
\eeq
From a generic multi-terminal consideration~\cite{PhysRevB.46.12485},
one can obtain several statements such as positive autocorrelation and negative cross-correlation for fermions at $T=0$
and possibility of a negative cross-correlation for bosons.
At the same time, it is important to specify precise forms of ${\cal T}$ and ${\cal L}$ for determination of behaviors of $S_I$ and $S_N$.
As specific examples relevant to cold-atom experiments, in what follows, we analyze QPC (Sec.~III) and  quantum dot systems (Sec.~IV).

We also note that  current fluctuations including the current noise can exhibit nontrivial frequency dependences even in the case of a non-driven noninteracting system.
At the same time, what is usually relevant for physics applications is the noises at the zero frequency limit including Johnson-Nyquist and shot noises~\cite{blanter2000,kobayashi2021shot} 
(note however~\cite{deblock2003detection}).
Therefore, for the rest of the paper, we wish to focus on noise properties at zero frequency.

\section{Quantum point contact system with a localized single-particle loss}
Inspired by the recent experiment~\cite{PhysRevA.100.053605}, we now analyze
the single-channel QPC system in which
the constriction between two reservoirs has a one-dimensional quantum wire structure.

The experiment~\cite{PhysRevA.100.053605} has realized the limit of perfect transmittance in the QPC
in which the conductance is quantized in units of $1/h$ without dissipation,
and a near-resonance optical tweezer applied in the constriction  induces
 a single-particle loss in a local manner.
In ultracold atomic gases, dynamics in the presence of the single-particle loss is known to be well-described by the following quantum master equation~\cite{breuer2002}:
\beq
\partial_{\tau}\hat{\rho}=i[\hat{\rho},\hat{H}]+\int dx\gamma(x)\Big[\hat{\psi}(x)\hat{\rho}\hat{\psi}^{\dagger}(x)-\frac{\{\hat{\psi}^{\dagger}(x)\hat{\psi}(x),\hat{\rho}\}}{2}\Big],\nonumber\\
\label{eq:lindblad}
\eeq
where $\gamma$ represents the dissipation 
strength~\footnote{Just in case, we point out that the usage of the Lindblad  equation is not mandatory,
and one can adopt any system satisfying Eqs.~\eqref{eq:transmittance},~\eqref{eq:loss-prob}, and~\eqref{eq:vacuum} for the lossy mesoscopic system.
The discussions in this section is thus designed to make one aware of applications in ultracold atomic gases.
}. 
The above equation is also called the Gorini-Kossakowski-Sudarshan-Lindblad equation
or simply Lindblad equation.

We turn to determine the $S$-matrix of such a lossy quantum point contact.
By using the Keldysh formalism that gives a simple expression on the above master equation~\cite{sieberer2016}, 
retarded Green's function in the one-dimensional wire $G^R$  obeys the following Dyson equation~\cite{PhysRevLett.122.040402,PhysRevB.101.144301,PhysRevA.106.053320}:
\beq
&&G^R(x,y,\omega)=G^R_0(x,y,\omega)\nonumber\\
&&\ \ +\int dx_1G^R_0(x,x_1,\omega) V(x_1)G^R(x_1,y,\omega),
\label{eq:dyson}
\eeq 
where 
\beq
V(x)=-i\gamma(x),
\eeq
and $G^R_0$ represents unperturbed retarded Green's function.
The explicit form of $G^R_0$ is given by~\cite{bruus2004many}
\beq
G_0^R(x,y,\omega)=-i\sqrt{\frac{m}{2\omega}}e^{i\sqrt{2m\omega} |x-y|},
\eeq
with $\omega>0$ and mass of a particle $m$.
A key insight is that Eq.~\eqref{eq:dyson} is equivalent to the Lippmann-Schwinger equation of retarded Green's function in the presence of the complex potential $V$.
Thus, it follows that the corresponding scattering state $\psi$ obeys~\cite{ryndyk2016theory}
\beq
\psi(x)=\psi^{\text{in}}(x)+\int dx_1G^R_0(x,x_1)V(x_1)\psi(x_1),
\label{eq:field}
\eeq
with the incoming state $\psi^{\text{in}}$. 

\begin{figure}[htbp]
\centering
\includegraphics[width=6.7cm]{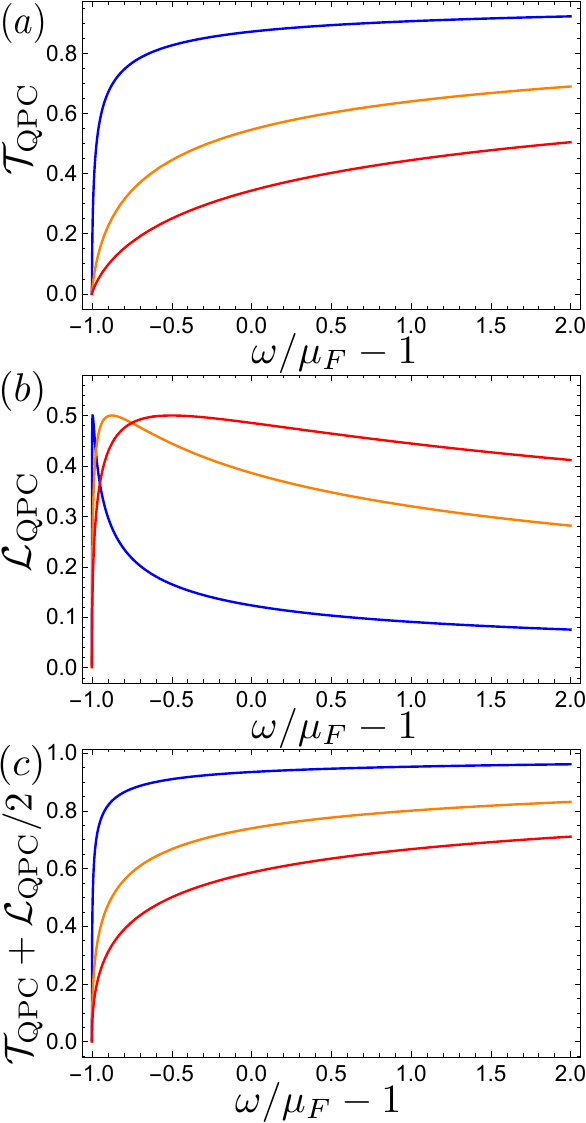}
\caption{\label{fig2}
Frequency dependences of transmittance (a) and loss probability (b) in the lossy QPC system.
The current between 1 and 2 reservoirs contains ${\cal T}_{\text{QPC}}+\frac{{\cal L}_{\text{QPC}}}{2}$ whose frequency dependence is shown in (c).
The blue, orange, and red curves are data with $\gamma\sqrt{m/\mu_F}=0.1$, $0.5$, and $1.0$, respectively.} 
\end{figure}

In what follows, we focus on
the simplest situation that the dissipation is local such  that $V(x)=-i\gamma\delta(x)$.
In order to determine the $S$-matrix,
we then consider the situation that the incoming state $\psi^{\text{in}}(x)=e^{i\sqrt{2m\omega}x}$ is scattered by $V$.
The  boundary conditions at infinity imply that $\psi(x)$ has the following form:
\beq
\psi(x)=\begin{cases}
e^{i\sqrt{2m\omega}x}+re^{-i\sqrt{2m\omega}x} \ \ (x<0)\\
te^{i\sqrt{2m\omega}x} \ \ \ \  \ \ \ \ \ \ \ \ \ \ \ \ \ \  \ (x>0)
\end{cases}
\label{eq:boundary1}
\eeq
By substituting \eqref{eq:boundary1} into \eqref{eq:field},
we obtain
\beq
r=\frac{-\gamma}{\sqrt{2\omega/m}+\gamma},\\
t=\frac{\sqrt{2\omega/m}}{\sqrt{2\omega/m}+\gamma}.
\eeq
We note that the result above is consistent with one in Refs.~\cite{PhysRevLett.122.040402,PhysRevB.101.144301}.
Similarly, for the incoming state $\psi^{\text{in}}(x)=e^{-i\sqrt{2m\omega}x}$,
one can consider the following solution:
\beq
\psi(x)=\begin{cases}
e^{-i\sqrt{2m\omega}x}+r'e^{i\sqrt{2m\omega}x} \ \ (x>0)\\
t'e^{-i\sqrt{2m\omega}x} \ \ \ \  \ \ \ \ \ \ \ \ \ \ \ \ \ (x<0)
\end{cases}
\label{eq:boundary2}
\eeq
By using above, we obtain
\beq
r'=r\\
t'=t.
\eeq
We notice that in contrast to the case of a real potential barrier, we have
\beq
|r|^2+|t|^2\ne1
\eeq
in the presence of $\gamma$.
By means of the unitary condition, the $S$-matrix is eventually obtained as
 \beq
 \hat{s}=\begin{pmatrix}
 r & t & le^{i\theta_1} \\
t &r& le^{i\theta_1} \\
le^{i\theta_2} & le^{i\theta_2}  &  
-(r+t) e^{i(\theta_1+\theta_2)}
 \end{pmatrix},
 \label{eq:s-matrix}
 \eeq
where $l= \frac{\sqrt{2\gamma}(2\omega/m)^{\frac{1}{4}}}{\sqrt{2\omega/m}+\gamma}$
$\theta_1$ and $\theta_2$ are phase factors that cannot be identified from the unitarity condition alone.
We point out that when $\theta_1=\theta_2=0$, the $S$-matrix corresponds to one introduced in the context of a three-terminal system~\cite{PhysRevA.30.1982}.
From Eqs.~\eqref{eq:transmittance} and~\eqref{eq:loss-prob}, we obtain
\beq
{\cal T}_{\text{QPC}}(\omega)=\frac{2\omega\theta(\omega)/m}{(\sqrt{2\omega/m}+\gamma)^2},\\
{\cal L}_{\text{QPC}}(\omega)=\frac{2\gamma\sqrt{2\omega/m}\theta(\omega)}{(\sqrt{2\omega/m}+\gamma)^2},
\eeq
whose frequency dependences for free fermions in which plots in units of the Fermi energy $\mu_F$ are allowed are shown in Fig.~\ref{fig2}.

Now that the $S$-matrix is determined, we turn to analyze the corresponding currents and noises.

When it comes to the current fluctuations at zero frequency, it is convenient to employ the method of  the full counting statistics.
This method was initially introduced in the context of photon counting in optics, 
and later on has been applied in the context of mesoscopic transport~\cite{levitov1993charge,levitov1996electron,nazarov2012quantum}.
The basic idea is that by constructing the generating function for emitted particles in each reservoir, one can 
obtain arbitrary order of the current fluctuations including the average current and noise.
For QPC systems, one can harness the so-called Levitov-Lesovik formula of the characteristic function~\cite{levitov1993charge,levitov1996electron,nazarov2012quantum}.
In the case of the three-terminal situation, it corresponds to the following characteristic function:
\beq
Z_{\text{QPC}}(\chi_1,\chi_2)=\det\Big[\hat{1}\pm\hat{n}(\hat{1}-\hat{s}^{\dagger}_{-\chi}\hat{s}_{\chi})\Big]^{\mp1},
\eeq
with
\beq
\hat{n}=\begin{pmatrix}
n_1 & 0 & 0\\
0 & n_2 & 0\\
0 & 0 & n_3
\end{pmatrix},
\eeq
\beq
\hat{s}_{\chi}=\begin{pmatrix}
r & t e^{-i\frac{(\chi_1-\chi_2)}{2}}& le^{i\theta_1} e^{-i\frac{\chi_1}{2}}\\
t e^{i\frac{(\chi_1-\chi_2)}{2}} & r & le^{i\theta_1}e^{-i\frac{\chi_2}{2}}\\
le^{i\theta_2}e^{i\frac{\chi_1}{2}} & le^{i\theta_2}e^{i\frac{\chi_2}{2}} & -(t+r)e^{i(\theta_1+\theta_2)}
\end{pmatrix}.
\eeq
Here, $\chi_1$ and $\chi_2$ are the counting fields to generate $I_1$ and $I_2$, respectively.
The above determinant is taken over $3\times3$ reservoirs and frequency spaces.
If one considers the statistics of the particle transfer among the time duration $\tau_0$,
the frequency space is quantized in units of $2\pi/\tau_0$~\cite{kamenev2011}.
With this understanding and Eq.~\eqref{eq:vacuum},
the characteristic function is obtained as
\begin{flalign}
Z_{\text{QPC}}(\chi_1,\chi_2)=\prod_{\omega}
\Big(1\pm{\cal T}_{\text{QPC}}\{(1\pm n_1)n_2(1-e^{-i\chi})\nonumber\\
+(1\pm n_2)n_1(1-e^{i\chi})\} \nonumber\\
\pm{\cal L}_{\text{QPC}}\{n_1(1-e^{i\chi_1})+n_2(1-e^{i\chi_2}) \}\Big)^{\mp1},
\end{flalign}
where $\chi=\chi_1-\chi_2$.
In order to generate the current correlation functions at arbitrary order, 
it is convenient to consider the cumulant generating function given by
\beq
F_{\text{QPC}}(\chi_1,\chi_2)=\frac{\log Z_{\text{QPC}}(\chi_1,\chi_2 )}{\tau_0}.
\eeq
By taking the large $\tau_0$ limit that allows to replace the sum over frequency by the corresponding integral, the cumulant generating function is obtained as
\begin{flalign}
F_{\text{QPC}}(\chi_1,\chi_2)=\mp\int \frac{d\omega}{2\pi}\log\Big(1\pm{\cal T}_{\text{QPC}}\{(1\pm n_1)n_2(1-e^{-i\chi})\nonumber\\
+(1\pm n_2)n_1(1-e^{i\chi})\} \nonumber\\
\pm{\cal L}_{\text{QPC}}\{n_1(1-e^{i\chi_1})+n_2(1-e^{i\chi_2}) \}\Big).
\label{eq:cumulant1}
\end{flalign}

\begin{figure}[htbp]
\centering
\includegraphics[width=6.7cm]{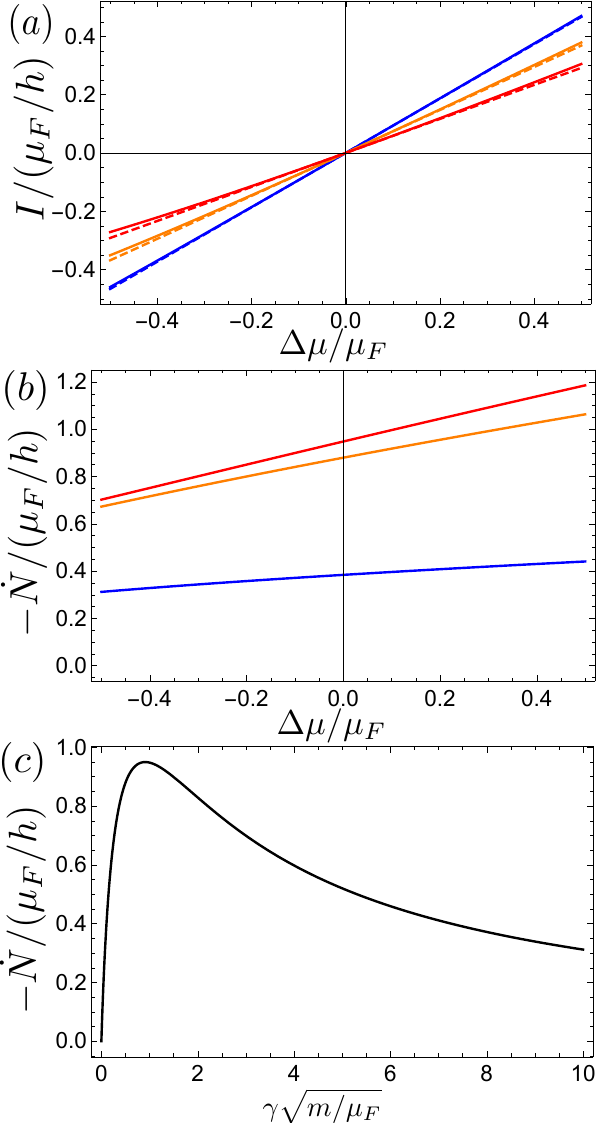}
\caption{\label{fig3}
(a) The (average) current-(chemical potential) bias characteristics in the presence of $\gamma$.
The blue, orange, and red curves are data with $\gamma\sqrt{m/\mu_F}=0.1$, $0.5$, and $1.0$, respectively.
The dashed lines are results with $I=G\Delta\mu$.
(b) The particle loss rate as a function of $\Delta\mu$. The different colors mean the results with different dissipation whose strengths are common to (a).
(c) The particle loss rate as a function of $\gamma$.
In all the plots, $T/\mu_F$ is chosen as 0.1.} 
\end{figure}
By using the above cumulant generating function, 
the average current and particle loss rate are respectively generated as follows:
\begin{flalign}
&I=\frac{1}{2}\Big[\frac{\partial F_{\text{QPC}}}{\partial (i\chi_1)}-\frac{\partial F_{\text{QPC}}}{\partial (i\chi_2)} \Big]\nonumber\\
&=\int \frac{d\omega}{2\pi}\Big[{\cal T}_{\text{QPC}}(\omega)+\frac{{\cal L}_{\text{QPC}}(\omega)}{2} \Big][n_1(\omega)-n_2(\omega)],
\end{flalign}
\beq
&-\dot{N}=\frac{1}{2}\Big[\frac{\partial F_{\text{QPC}}}{\partial (i\chi_1)}+\frac{\partial F_{\text{QPC}}}{\partial (i\chi_2)} \Big]\nonumber\\
&\ \ \ \ \ \ \ \ \ \ \ \ =\int \frac{d\omega}{2\pi}{\cal L}_{\text{QPC}}(\omega)[n_1(\omega)+n_2(\omega)].
\eeq
Of course, these are consistent with the results in the previous section.

In addition, the current noise is generated as follows:
\beq
S_{ij}(\omega=0)\equiv S_{ij}=\frac{\partial^2F_{\text{QPC}}}{\partial(i\chi_i)\partial(i\chi_j)}.
\eeq
Thus, we obtain
\begin{widetext}
\beq
S_{11}=\int\frac{d\omega}{2\pi}\Big[{\cal T}_{\text{QPC}}(n_1+n_2\pm2n_1n_2)+{\cal L}_{\text{QPC}}n_1\pm\{{\cal T}_{\text{QPC}}(n_1-n_2)+{\cal L}_{\text{QPC}}n_1\}^2\Big],
\eeq
\beq
S_{22}=\int\frac{d\omega}{2\pi}\Big[{\cal T}_{\text{QPC}}(n_1+n_2\pm2n_1n_2)+{\cal L}_{\text{QPC}}n_2\pm\{{\cal T}_{\text{QPC}}(n_2-n_1)+{\cal L}_{\text{QPC}}n_2\}^2\Big],
\eeq
\beq
S_{12}=S_{21}=\int\frac{d\omega}{2\pi}\Big[-{\cal T}_{\text{QPC}}(n_1+n_2\pm2n_1n_2)\pm\{ {\cal T}_{\text{QPC}}(n_2-n_1)-{\cal L}_{\text{QPC}}n_1 \}\{{\cal T}_{\text{QPC}}(n_1-n_2)-{\cal L}_{\text{QPC}}n_2  \}\Big].
\eeq
It is straightforward to confirm that each element above is consistent with Eq.~\eqref{eq:generic-noise}.
By using above, the current noise between 1 and 2 reservoirs is given by
\beq
S_I=\int\frac{d\omega}{2\pi}\Big[{\cal T}_{\text{QPC}}\{n_1(1\pm n_1)+n_2(1\pm n_2)\}+\frac{{\cal L}_{\text{QPC}}}{4}(n_1+n_2)
\pm\Big\{{\cal T}_{\text{QPC}}({\cal T}_{\text{QPC}}-1 )+{\cal T_{\text{QPC}}L_{\text{QPC}}}+\frac{{\cal L}^2_{\text{QPC}}}{4}\Big\}(n_1-n_2)^2
\Big].\nonumber\\
\label{eq:current-noise-QPC}
\eeq
\end{widetext}
We note that in the absence of dissipation where ${\cal L}_{\text{QPC}}=0$, the above reduces to one known in two-terminal systems~\cite{blanter2000}.
In this particular case, the first term  on the right hand side is sometimes called equilibrium noise contribution and
the last term that changes sign depending on quantum statistics is the so-called shot noise contribution.
Indeed, in the zero bias limit in which $n_1=n_2\equiv n$, we have
\beq
\int\frac{d\omega}{2\pi}{\cal T}_{\text{QPC}}\{n_1(1\pm n_1)+n_2(1\pm n_2)\}\to2TG.
\label{eq:jn-noise}
\eeq
Here, we use $(1\pm n)n=-T\frac{\partial n}{\partial\omega}$ and introduce the conductance
\beq
G=-\int\frac{d\omega}{2\pi}{\cal T}_{\text{QPC}}\frac{\partial n}{\partial\omega}.
\eeq
Equation~\eqref{eq:jn-noise} is the familiar Johnson-Nyquist noise expected at equilibrium.
In contrast, the significance of the shot noise contribution becomes clearer for free fermions at $T=0$.
If the frequency dependence in ${\cal T}_{\text{QPC}}$ is negligible~\footnote{In our model, ${\cal T}_{\text{QPC}}=1$,  regardless of frequency in the absence of dissipation.}, we have
\beq
-\int \frac{d\omega}{2\pi} {\cal T}_{\text{QPC}}({\cal T}_{\text{QPC}}-1)(n_1-n_2)^2\nonumber\\
\to\frac{|\Delta\mu|}{2\pi} {\cal T}_{\text{QPC}}(1-{\cal T}_{\text{QPC}}),
\eeq
which is the famous shot noise formula for two-terminal systems~\cite{blanter2000}. Moreover, hypothetically assuming that the channel transmittance is small such that 
${\cal T}_{\text{QPC}}(1-{\cal T}_{\text{QPC}})\approx{\cal T}_{\text{QPC}}$,
the above reduces to the Schottky formula noticing that the ratio of the current noise to the average current is given by a charge of a 
particle~\footnote{For electron systems, an electric current $I_e=eI$ is usually concerned and the Schottky formula is then given by $S_{I_e}/I_e=e$.
In the case of charge neutral ultracold atomic gases, a particle current is concerned and thereby the Schottky formula is given by $S_I/I=1$.}.

\begin{figure}[htbp]
\centering
\includegraphics[width=6.7cm]{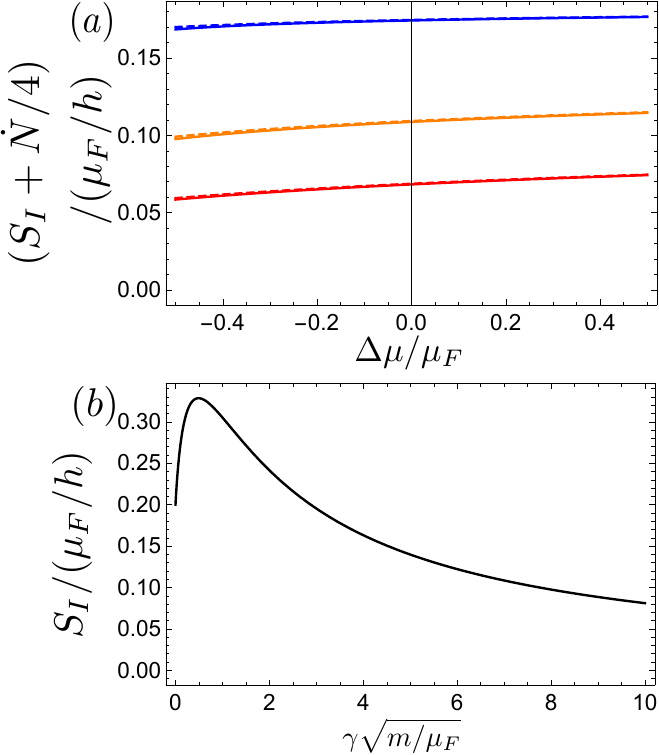}
\caption{\label{fig4}
(a) Behaviors of $S_I+\frac{\dot{N}}{4}$ as a function of $\Delta\mu$ for different dissipation strengths.
The blue, orange, and red colors mean data with $\gamma\sqrt{m/\mu_F}=0.1$, $0.5$, and $1.0$, respectively.
The dashed curves are results with Eq.~\eqref{eq:approx-noise}.
(b) The current noise at $\Delta\mu=0$ as a function of $\gamma$. 
In all the plots, $T/\mu_F=0.1.$} 
\end{figure}
In the presence of dissipation, the second term proportional to the particle loss rate is present 
and the shot noise contribution is modified such that terms proportional to ${\cal T}_{\text{QPC}}{\cal L}_{\text{QPC}} $ and ${\cal L}^2_{\text{QPC}}$ are present
in integrand. From the explicit forms of ${\cal T}_{\text{QPC}} $ and ${\cal L}_{\text{QPC}} $, however, one can confirm the following identity:
\beq
{\cal T}_{\text{QPC}}({\cal T}_{\text{QPC}}-1 )+{\cal T_{\text{QPC}}L_{\text{QPC}}}+\frac{{\cal L}^2_{\text{QPC}}}{4}=0.
\label{eq:identity}
\eeq
As a consequence, the current noise in the lossy QPC system reduces to
\beq
S_I=-\frac{\dot{N}}{4}-T
\int\frac{d\omega}{2\pi}{\cal T}_{\text{QPC}}\Big[\frac{\partial n_1}{\partial\omega}+\frac{\partial n_2}{\partial\omega}\Big].
\eeq
This result shows that the current noise between 1 and 2 reservoirs is expressed with the sum of the particle loss rate and equilibrium noise contribution.
What is remarkable here is that the shot noise contribution is absent although the channel transmittance decreases from 1 due to dissipation.

\begin{figure}[htbp]
\centering
\includegraphics[width=6.7cm]{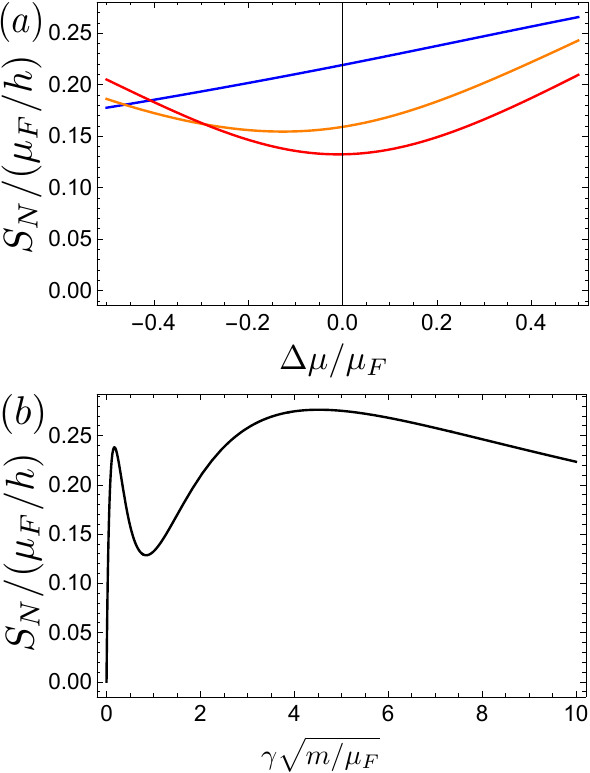}
\caption{\label{fig5}
(a) The noise of the particle loss rate as a function of $\Delta\mu$.
The blue, orange, and red curves are data with $\gamma\sqrt{m/\mu_F}=0.1$, $0.5$, and $1.0$, respectively.
(b) $S_N$ at $\Delta\mu=0$ as a function of $\gamma$.
In all the plots, $T/\mu_F=0.1$. } 
\end{figure}
Finally, the noise of the particle loss rate is rewritten as
\beq
S_N=-\dot{N}\pm\int\frac{d\omega}{2\pi}{\cal L}^2_{\text{QPC}}(n_1+n_2)^2.
\label{eq:noise-loss-QPC}
\eeq
Thus, this noise consists of the particle loss rate and the contribution that changes sign depending on bosons or fermions.
As both contributions contain the factor $n_1+n_2$, 
the integrands take nonzero values over wide ranges of frequencies.
Moreover, the presence of the factor ${\cal L}_{\text{QPC}}$ leads to the absence of $S_N$ for $\gamma=0$.

We now apply the transport quantities calculated above to the free fermion case, which is directly relevant to the experiment in Ref.~\cite{PhysRevA.100.053605}.

Figure~\ref{fig3} shows the average current and particle loss rate for different dissipation strengths.
As can be seen from Fig.~\ref{fig3}~(a), at a fixed chemical potential bias, the average current monotonically decreases with increasing $\gamma$.
It is also clear that regardless of dissipation strengths there is a small bias regime in which the average current is proportional to $\Delta\mu$, i.e., Ohmic regime.
In contrast, as shown in Fig.~\ref{fig3}~(b), the particle loss rate becomes nonzero regardless of an applied bias once a nonzero $\gamma$ is introduced.
This is due to the fact that the integrand of the particle loss rate contains the sum of the distribution functions of two terminals, 
whilst the integrand of the average current contains the difference between the distribution functions and therefore 
the average current vanishes at zero bias.
Although the particle loss appears to be monotonically enhanced with increasing $\gamma$, it does not.
As shown in~Fig.~\ref{fig3}~(c), beyond certain threshold, the particle loss is rather suppressed with increasing $\gamma$.
This suppression behavior can be understood from the fact that the integrand of the particle loss rate contains ${\cal L}_{\text{QPC}}$,
which increases with $\gamma$ below the threshold value and then decreases with $\gamma$ beyond it.

Figure~\ref{fig4} shows the current noise between 1 and 2 reservoirs.
As mentioned before, in the absence of dissipation, the current noise is purely determined from the so-called equilibrium noise
contribution that is equivalent to the Johnson-Nyquist noise in the zero bias limit.
Since the integrand of this contribution contains ${\cal T}_{\text{QPC}}$, 
the equilibrium noise contribution is monotonically suppressed as increasing $\gamma$.
When it comes to a small bias regime, one can obtain the following approximate expression:
\beq
S_I+\frac{\dot{N}}{4}\approx\frac{T[{\cal T}_{\text{QPC}}(\mu_F)+{\cal T}_{\text{QPC}}(\mu_F+\Delta\mu)]}{2\pi}.
\label{eq:approx-noise}
\eeq
In order to obtain above, we use the fact that ${\cal T}_{\text{QPC}}$ is the smooth function near the Fermi level and
use the following formula:
\begin{flalign}
\int dx [1-\tanh(x-y)\tanh(x+y)]\nonumber\\
=\coth(2y)\log\Big(\frac{\cosh(x+y)}{\cosh(x-y)}\Big).
\label{eq:integral-formula}
\end{flalign}
As shown in Fig.~\ref{fig4}~(a), the above approximate formula gives excellent agreements with the exact analysis even at $T/\mu_F=0.1$.
In addition, once dissipation is present, the particle loss rate contribution gives rise to 
a non-negligible impact, and correspondingly the total current noise shows the non-monotonic behavior shown in Fig.~\ref{fig4}~(b).

A different monotonicity is present in the noise of the particle loss rate, which is shown in Fig.~\ref{fig5}.
Since the first and second terms in Eq.~\eqref{eq:noise-loss-QPC} exhibit different peak structures and they have opposite sign, 
  $S_N$ is found to manifect the double peak structure as a function of $\gamma$ as shown in Fig.~\ref{fig5}~(b).
 
\section{Keldysh study of a  lossy quantum dot}
As a complementary single-channel mesoscopic system, we now consider the noninteracting quantum dot system coupled to
two macroscopic reservoirs. Whilest a single-channel quantum dot system has yet to be realized in ultracold atomic gases,
it is potentially reachable by means of a digital mirror device~\cite{PhysRevLett.119.030403}.
We also note that the theoretical consideration to the corresponding multi-site systems discussed in Ref.~\cite{PhysRevA.106.053320} is straightforward.

As in the case of the previous section, we first obtain the cumulant generating function.
To this end, it is convenient to adopt the so-called two-point measurement scheme in the full counting statistics~\cite{PhysRevB.78.115429,RevModPhys.81.1665}.
The basic quantity in the two-point measurement is the joint probability to measure the particle number in each reservoir at initial time 0 and at time $\tau_0$.
The probability distribution of particles transferred from each reservoir to the conduction channel during the interval $\tau_0$ is obtained by taking into account
every initial and final conditions on the transition probability. The characteristic function is then given by the Fourier transformation of this probability distribution and 
is expressed as~\cite{PhysRevB.78.115429,RevModPhys.81.1665}
\beq
Z(\chi_1,\chi_2)=\langle\hat{U}^{\dagger}_{-\chi}(\tau_0,0)\hat{U}_{\chi}(\tau_0,0)\rangle,
\eeq
where
\beq
\hat{U}_{\chi}(\tau_0,0)=e^{-i\int_0^{\tau_0}d\tau [e^{\frac{i\chi_1\hat{N}_1+i\chi_2\hat{N}_2}{2}}\hat{H}e^{\frac{-i\chi_1\hat{N}_1-i\chi_2\hat{N}_2}{2}}] },
\eeq
with a Hamiltonian $H$, and  the average above is taken with the density matrix at the initial time $\hat{\rho}(0)$ in which there is no correlation between reservoirs and conduction channel.
We note that the result above is obtained for closed (total) systems  including both two-terminal system and third reservoir absorbing particles.

In order to tune the above result to the lossy system whose dynamics is governed by the Lindblad equation~\eqref{eq:lindblad},
here we employ the method with stochastic fields, which has been used in the context of photon counting~\cite{zoller1997quantum},
non-hermitian response theory~\cite{pan2020non}, and transport with dephasing~\cite{PhysRevB.102.100301,PhysRevResearch.4.013109}.
In the presence of a single-particle loss, one can consider the following stochastic Hamiltonian:
 \beq
 \hat{H}_{\eta}=\hat{H}+\hat{V}_{\eta},
 \eeq
where $\hat{H}$ is the two-terminal quantum dot Hamiltonian given by
\beq
&&\hat{H}=\hat{H}_1+\hat{H}_2+\hat{H}_T+\hat{H}_{d},\\
&&\hat{H}_T=\sum_{j=1,2}\sum_{\mk}t_{\mk}\hat{\psi}^{\dagger}_{j,\mk}\hat{d}+\text{h.c.},\\
&&\hat{H}_d=\epsilon \hat{d}^{\dagger}\hat{d},
\eeq
with reservoir Hamiltonians $\hat{H}_1$ and $\hat{H}_2$, and energy of the quantum dot $\epsilon$.
In addition, $\hat{V}_{\eta}$ represents the interaction between quantum dot and stochastic fields given by
\beq
\hat{V}_{\eta}=\gamma[\hat{d}^{\dagger}\hat{\eta}+\hat{\eta}^{\dagger} \hat{d}].
\eeq
According to the stochastic Hamiltonian above, one can define the corresponding density matrix $\hat{\rho}_{\eta}$ and characteristic function $Z(\chi_1,\chi_2,\eta)$ for a given noise realization.
In order to reproduce dynamics obeying the Lindblad equation~\eqref{eq:lindblad}, it is necessary to impose the following noise averages:
\beq
&&\langle \hat{\eta}(\tau)\hat{\eta}^{\dagger}(\tau')\rangle_{\eta}=\delta(\tau-\tau'),\label{eq:noise1}\\
&&\langle \hat{\eta}^{\dagger}(\tau)\hat{\eta}(\tau')\rangle_{\eta}=\langle \hat{\eta}(\tau)\hat{\eta}(\tau')\rangle_{\eta}=0. \label{eq:noise2}
\eeq

Based on the above preparation, we turn to calculate the characteristic function. 
As it is expressed with forward and backward time evolution operators, it is convenient to adopt the Keldysh field theory approach~\cite{kamenev2011}.
Since the retarded, advanced, and Keldysh components in Green's functions of the noise fields are respectively expressed as
\beq
g^R_{\eta}(\tau)=-i\theta(\tau)\langle[\hat{\eta}(\tau),\hat{\eta}^{\dagger}(0)]_{\mp}\rangle_{\eta},\\
g^A_{\eta}(\tau)=i\theta(-\tau)\langle[\hat{\eta}(\tau),\hat{\eta}^{\dagger}(0)]_{\mp}\rangle_{\eta},\\
g^K_{\eta}(\tau)=-i\langle[\hat{\eta}(\tau),\hat{\eta}^{\dagger}(0)]_{\pm}\rangle_{\eta},
\eeq
their Fourier components are obtained as
\beq
g^R_{\eta}(\omega)=-\frac{i}{2},\\
g^A_{\eta}(\omega)=\frac{i}{2}\\
g^K_{\eta}(\omega)=-i.
\eeq
Here, we adopt the proper convention of the Heviside step function in the Keldysh analysis~\cite{PhysRevB.102.100301}:
\beq
\theta(0)=\frac{1}{2}.
\eeq
Due to the presence of $V_{\eta}$, Green's functions in the dot are modified by the noise fields and
the average effects are accumulated in the self-energy $\Sigma$ as follows:
\beq
\Sigma^{R}(\omega)=-\frac{i\gamma}{2},\\
\Sigma^{A}(\omega)=\frac{i\gamma}{2},\\
\Sigma^{K}(\omega)=-i\gamma,
\eeq
where the superscripts $R$, $A$, and $K$ represent the retarded, advanced, and Keldysh components of the self-energy.
We note that the self-energy above is indeed consistent with the analysis based on the Lindblad equation~\cite{PhysRevA.106.053320}.
It is convenient to employ the path-integral representation of the Keldysh formalism for the characteristic function~\cite{kamenev2011}.
By integrating out the noise degrees of freedom and using
\beq
&& [\hat{N}_j,\hat{H}_{\eta}]=[\hat{N}_j,\hat{H}_T],\\
&& e^{\mp i\chi \hat{N}_l}\hat{H}_Te^{\pm i\chi \hat{N}_l}=\sum_{j=1,2}\sum_{\mk}t_{\mk}e^{\mp i \delta_{j,l} \chi}\hat{\psi}^{\dagger}_{j,\mk}\hat{d}+\text{h.c.}\nonumber\\
\eeq
the noise-averaged characteristic function  is  obtained as
\beq
Z_{\text{dot}}(\chi_1,\chi_2)=\int {\cal D}[\bar{\psi},\psi,\bar{d},d]e^{iS_1+S_2+iS_T+iS_d}.
\eeq
Here, the action part is expressed as follows:
\beq
S_1+S_2+S_T+S_d=\tau_0\sum_{\mk}\int \frac{d\omega}{2\pi}\Big[\bar{\Psi}_1\mathbf{g}^{-1}_1\Psi_1 \nonumber\\
+\bar{\Psi}_2\mathbf{g}^{-1}_2\Psi_2+\bar{\Psi}_1\mathbf{T}_{\chi_1}\mathbf{D}+\bar{\mathbf{D}}\mathbf{T}^{\dagger}_{\chi_1}\Psi_1 \nonumber\\
+\bar{\Psi}_2\mathbf{T}_{\chi_2}\mathbf{D}+\bar{\mathbf{D}}\mathbf{T}^{\dagger}_{\chi_2}\Psi_2 +\bar{\mathbf{D}}\mathbf{g}_d^{-1}\mathbf{D}
\Big].
\eeq
In the Keldysh formalism, we consider the closed time contour consisting of forward and backward branches, and correspondingly,
define fields $\psi^{+(-)}$, $d^{+(-)}$ residing on the forward (backward) branches.
For convenient, we introduce a new pair of fields whose vector representation is given by
\beq
\Psi=\begin{pmatrix}
\frac{\psi^{+}+\psi^{-}}{\sqrt{2}}\\
\frac{\psi^{+}-\psi^{-}}{\sqrt{2}}\
\end{pmatrix}, \ \
\mathbf{D}=\begin{pmatrix}
\frac{d^{+}+d^{-}}{\sqrt{2}}\\\
\frac{d^{+}-d^{-}}{\sqrt{2}}\
\end{pmatrix}.
\eeq
\begin{figure}[htbp]
\centering
\includegraphics[width=6.7cm]{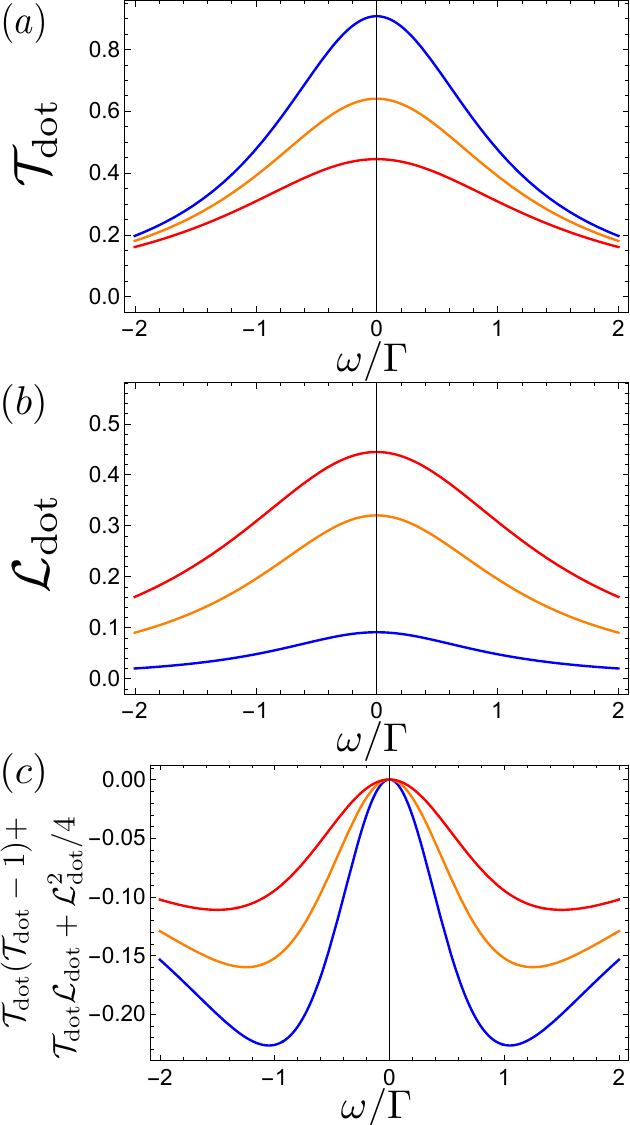}
\caption{\label{fig6}
Transmittance (a) and loss probability (b) as a function of $\omega$ in the lossy quantum dot system.
The energy level of the dot $\epsilon$ is set as 0.
In contrast to the lossy quantum point contact system, ${\cal T}_{\text{dot}}({\cal T}_{\text{dot}}-1)+{\cal T}_{\text{dot}}{\cal L}_{\text{dot}}+{\cal L}^2_{\text{dot}}/4 $ becomes nonzero as shown in (c).
The blue, orange, and red curves are data with $\gamma/\Gamma=0.1$, $0.5$, and $1.0$, respectively.} 
\end{figure}
Correspondingly, inverse of Green's functions and tunneling amplitudes can be expressed with $2\times2$ matrices as follows:
\beq
\mathbf{g}^{-1}=\begin{pmatrix}
0 & [g^{-1}(\mk,\omega)]^{A}\\
 [g^{-1}(\mk,\omega)]^{R} & [g^{-1}(\mk,\omega)]^{K}
\end{pmatrix}
\nonumber\\=\begin{pmatrix}
0 & \omega-\xi_k-i0^+\\
\omega-\xi_k+i0^+ & 2i0^+[1\pm 2n(\omega)]
\end{pmatrix},
\eeq
\beq
\mathbf{g}_d^{-1}=\begin{pmatrix}
0 & \omega-\epsilon-\frac{i\gamma}{2}\\
\omega-\epsilon+\frac{i\gamma}{2} & i\gamma
\end{pmatrix},
\eeq
\beq
\mathbf{T}_{\chi}=t_{\mk}\begin{pmatrix}
 -i\sin\frac{\chi}{2} & \cos\frac{\chi}{2}\\
 \cos\frac{\chi}{2} & -i\sin\frac{\chi}{2} 
\end{pmatrix}.
\eeq
We note that in the Keldysh field theory analysis 
the origin of energy (frequency) is taken as one at the chemical potential and therefore the distribution function appeared above 
should be read as $n(\omega)=\frac{1}{e^{\omega/T}\mp1 }$~\cite{PhysRevA.106.053320}. 
Since the action above is quadratic in fields, one can exactly perform the integration in the characteristic function.
By doing it, the cumulant generating function that is essentially logarithm of the characteristic function is given by
\beq
F_{\text{dot}}(\chi_1,\chi_2)&=&\mp\int \frac{d\omega}{2\pi}\log\Big(1\pm{\cal T}_{\text{dot}}\{(1\pm n_1)n_2(1-e^{-i\chi})\nonumber\\
&&+(1\pm n_2)n_1(1-e^{i\chi})\} \nonumber\\
&&\pm{\cal L}_{\text{dot}}\{n_1(1-e^{i\chi_1})+n_2(1-e^{i\chi_2}) \}\Big),
\label{eq:cumulant2}
\eeq
where
\beq
{\cal T}_{\text{dot}}(\omega)=\frac{[\Gamma(\omega)]^2}{(\omega-\epsilon-R(\omega))^2+(\Gamma(\omega)+\frac{\gamma}{2})^2},\\
{\cal L}_{\text{dot}}(\omega)=\frac{\gamma\Gamma(\omega)}{(\omega-\epsilon-R(\omega))^2+(\Gamma(\omega)+\frac{\gamma}{2})^2},
\eeq
with
\beq
R(\omega)=2\text{Re}[\sum_{\mk}|t_{\mk}|^2g^R(\mk,\omega)],\\
\Gamma(\omega)=2\text{Im}[\sum_{\mk}|t_{\mk}|^2g^R(\mk,\omega)].
\eeq
We point out that the formal expressions of the characteristic functions between the QPC and quantum dot systems coincide in that 
Eq.~\eqref{eq:cumulant2} corresponds to Eq.~\eqref{eq:cumulant1} by replacements ${\cal T}_{\text{dot}}\to {\cal T}_{\text{QPC}}$ and ${\cal L}_{\text{dot}}\to {\cal L}_{\text{QPC}}$.

As in the case of the QPC, we now look at the free fermion case.
For the sake of simplicity, we analyze the system in the wide-band limit where
$R(\omega)\to0$ and
$\Gamma(\omega)\to\Gamma$,
and thereby the transport quantities can be analyzed in units of $\Gamma$.

Figure~\ref{fig6} shows channel transmittance and loss probability at $\epsilon=0$.
They are the Lorentzian functions peaked at $\epsilon$, which is in sharp contrast to the QPC case.

\begin{figure}[htbp]
\centering
\includegraphics[width=6.7cm]{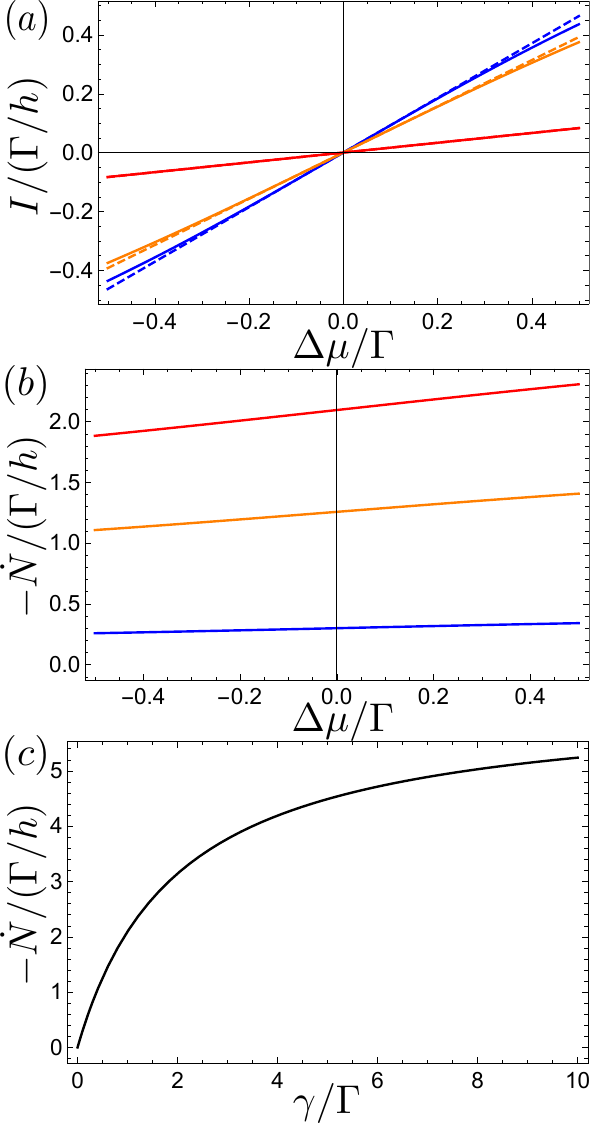}
\caption{\label{fig7}
The particle current (a) and particle loss rate (b) as a function of $\Delta\mu$ in the quantum dot system.
The dashed lines are results with $I=G\Delta\mu$.
The blue, orange, and red curves are data with $\gamma/\Gamma=0.1$, $0.5$, and $1.0$, respectively.
(c) The particle loss rate as a function of $\gamma$.
In all the plots, $T/\Gamma=0.1$ and $\epsilon=0$.} 
\end{figure}

\begin{figure}[htbp]
\centering
\includegraphics[width=6.7cm]{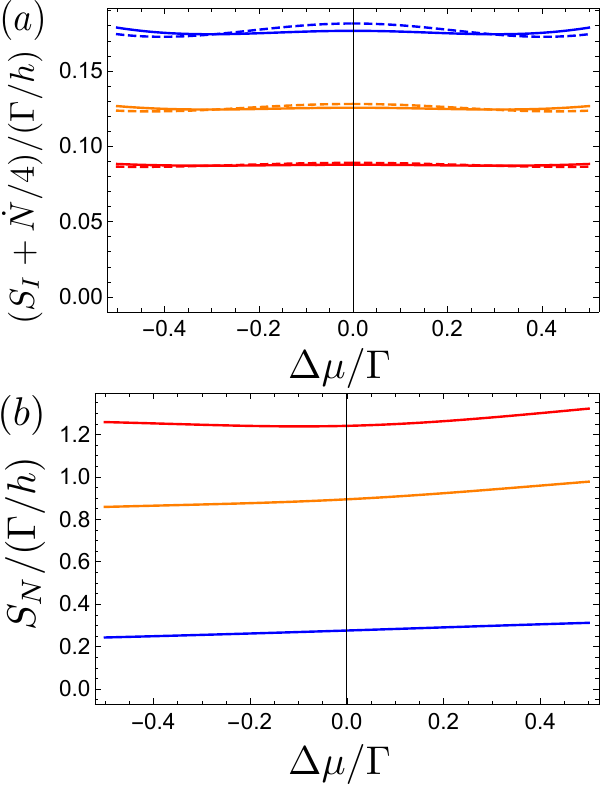}
\caption{\label{fig8}
(a) The current noise in which the particle loss rate contribution is subtracted as a function of $\Delta\mu$ in the quantum dot system.
The dashed curves are results with Eq.~\eqref{eq:approx-noise2}.
(b) The noise behavior of the particle loss rate.
The blue, orange, and red curves are data with $\gamma/\Gamma=0.1$, $0.5$, and $1.0$, respectively.
In all the plots, $T/\Gamma=0.1$ and $\epsilon=0.$} 
\end{figure}
Figure~\ref{fig7} shows the average current and particle loss rate by means of ${\cal T}_{\text{dot}}$ and ${\cal L}_{\text{dot}}$.
As in the case of the QPC,
the average current shows the Ohmic behavior at small chemical potential biases 
and its absolute value is weakened by the presence of dissipation.
In addition, the particle loss rate shows a small bias dependence.
What is different from the QPC situation is 
the monotonic increase of the particle loss rate as a function of $\gamma$, which can also be confirmed from Fig.~\ref{fig7}~(c).
Such a difference can emerge, since frequency integration over a wide range is relevant for evaluation of $-\dot{N}$.

We next look at $S_I$.
An important observation is  that in contrast to the QPC, the quantum dot system does not obey the identity \eqref{eq:identity}, i.e.,
\beq
{\cal T}_{\text{dot}}({\cal T}_{\text{dot}}-1 )+{\cal T_{\text{dot}}L_{\text{dot}}}+\frac{{\cal L}^2_{\text{dot}}}{4}\ne0 
\eeq 
except for $\omega=\epsilon$. This result implies that the shot noise contribution must be present.
Figure~\ref{fig8}~(a) shows the current noise as a function of $\Delta\mu$ in different dissipation strengths.
Although the dominant contribution arises from the particle loss rate contribution at moderate dissipation strengths, there are also nonzero
equilibrium noise and shot noise contributions.
In fact, except for the particle loss rate contribution, the current noise in free fermions can be expressed 
in terms of difference of distribution functions.
Thus, when it comes to the low-bias regime at low temperatures, 
we can obtain the following approximate expression:
\begin{flalign}
S_I+\frac{\dot{N}}{4}\approx\frac{T\{{\cal T}_{\text{dot}}^2+{\cal T}_{\text{dot}}{\cal L}_{\text{dot}}+{\cal L}_{\text{dot}}^2/4 \}_{\omega=0} }{2\pi}\nonumber\\
+\frac{T\{{\cal T}_{\text{dot}}^2+{\cal T}_{\text{dot}}{\cal L}_{\text{dot}}+{\cal L}_{\text{dot}}^2/4 \}_{\omega=\Delta\mu} }{2\pi}\nonumber\\
-\frac{\Delta\mu\coth(\frac{\Delta\mu}{2T}) \{{\cal T}_{\text{dot}}({\cal T}_{\text{dot}}-1 )+{\cal T}_{\text{dot}}{\cal L}_{\text{dot}}+{\cal L}_{\text{dot}}^2/4 \}_{\omega=\frac{\Delta\mu}{2}} }{2\pi}.
\label{eq:approx-noise2}
\end{flalign}
We note that Eq.~\eqref{eq:integral-formula} is applied to obtain the above result.
The comparisons between exact  and approximate expressions are shown in Fig.~\ref{fig8}~(a).
One may argue that compared with the QPC, there are non-negligible deviations between them.
We find that the deviations originate from a finite-temperature effect and
confirm the fact that the deviations become smaller at lower temperatures.

Finally, we discuss the noise of the particle loss rate. 
We note that  Eq.~\eqref{eq:noise-loss-QPC} is valid provided that the replacement ${\cal L}_{\text{QPC}}\to{\cal L}_{\text{dot}}$ is done.
Figure~\ref{fig8}~(b) shows $S_N$ for different dissipation strengths and indicates that
$S_N$ monotonically increases as increasing $\gamma$.
The absence of a non-monotonic behavior appeared in the QPC system
stems from the fact that the particle loss rate contribution in $S_N$ giving rise to the dominant effect 
is the monotonically increasing function in $\gamma$, although another contribution in $S_N$ shows a non-monotonic behavior in $\gamma$.

\section{summary}
We have examined the transport properties of the single-channel lossy mesosocpic systems
that include the average current, particle loss rate, current noise, and noise of the particle loss rate.
By analyzing the QPC and quantum dot systems,
we have revealed that the cumulant generating function of the current is common to two systems and
the noises contain the component proportional to the particle loss rate. 
Since the particle loss rate has already been measured in the lossy QPC system~\cite{PhysRevA.100.053605},
the current noise component that is the same order with the particle loss rate may be detectable in experiments.

Although this work focuses on non-interacting reservoirs,
our formulation can also be applied to fermionic superfluid reservoirs.
There, it is interesting to investigate how a particle loss affects the current noise involving effects of multiple Andreev reflections~\cite{nazarov2012quantum},
which  has already been measured under a dissipationless condition in condensed matter~\cite{PhysRevLett.86.4104}.
In addition, extensions to bosonic superfluids~\cite{PhysRevLett.116.235302,mullers2018coherent}
and synthetic dimensions~\cite{ono2021} are relevant in experiments of ultracold atomic gases and therefore
are promising future works.

\section*{acknowledgment}
The author thanks A. Aharony, T. Esslinger, P. Fabritius, T. Giamarchi, M.-Z. Huang, J. Mohan, M. Talebi, M. Ueda, A.-M. Visuri, and S. Wili
for useful discussions that motivate this work.
This work is supported by MEXT Leading Initiative for Excellent Young Researchers (Grant No.~JPMXS032020000),
JSPS KAKENHI (Grant No.~JP21K03436), and Matsuo Foundation.

%

\end{document}